\documentclass[twocolumn]{aastex631}

\usepackage{epsfig,graphics,subfigure,psfrag,amsmath,amssymb,amscd}

\begin{document}

\title{A Unified Timescale Relation for Quasi-Periodic Eruptions and Repeated Nuclear Transients}

\author[0000-0001-7689-6382]{Shifeng Huang}
\affiliation{
Department of Astronomy, University of Science and Technology of China, Hefei, 230026, China; sfhuang999@ustc.edu.cn;twang@ustc.edu.cn; jnac@ustc.edu.cn}
\affiliation{School of Astronomy and Space Sciences,
University of Science and Technology of China, Hefei, 230026, China}

\author[0000-0002-1517-6792]{Tinggui Wang}
\affiliation{ 
Department of Astronomy, University of Science and Technology of China, Hefei, 230026, China; sfhuang999@ustc.edu.cn;twang@ustc.edu.cn; jnac@ustc.edu.cn}
\affiliation{School of Astronomy and Space Sciences,
University of Science and Technology of China, Hefei, 230026, China}
\affiliation{Department of Physics and Astronomy, College of Physics, Guizhou University, Guiyang 550025, People's Republic of China}

\author[0000-0002-7152-3621]{Ning Jiang}
\affiliation{
Department of Astronomy, University of Science and Technology of China, Hefei, 230026, China; sfhuang999@ustc.edu.cn;twang@ustc.edu.cn; jnac@ustc.edu.cn}
\affiliation{School of Astronomy and Space Sciences,
University of Science and Technology of China, Hefei, 230026, China}

\author[0000-0003-4225-5442]{Yibo~Wang}
\affiliation{ 
Department of Astronomy, University of Science and Technology of China, Hefei, 230026, China; sfhuang999@ustc.edu.cn;twang@ustc.edu.cn; jnac@ustc.edu.cn}
\affiliation{School of Astronomy and Space Sciences,
University of Science and Technology of China, Hefei, 230026, China}

\author[0000-0001-6938-8670]{Zhenfeng~Sheng}
\affiliation{Institute of Deep Space Sciences, Deep Space Exploration Laboratory, Hefei, 230026, China}

\author[0000-0002-8438-8529]{Tian-Yu Xia}
\affiliation{
Department of Astronomy, University of Science and Technology of China, Hefei, 230026, China; sfhuang999@ustc.edu.cn;twang@ustc.edu.cn; jnac@ustc.edu.cn}
\affiliation{School of Astronomy and Space Sciences,
University of Science and Technology of China, Hefei, 230026, China}

\author[0000-0003-3824-9496]{Jiazheng Zhu}
\affiliation{ 
Department of Astronomy, University of Science and Technology of China, Hefei, 230026, China; sfhuang999@ustc.edu.cn;twang@ustc.edu.cn; jnac@ustc.edu.cn}
\affiliation{School of Astronomy and Space Sciences,
University of Science and Technology of China, Hefei, 230026, China}

\author[0000-0003-4959-1625]{Zheyu Lin}
\affiliation{ 
Department of Astronomy, University of Science and Technology of China, Hefei, 230026, China; sfhuang999@ustc.edu.cn;twang@ustc.edu.cn; jnac@ustc.edu.cn}
\affiliation{School of Astronomy and Space Sciences,
University of Science and Technology of China, Hefei, 230026, China}

\author[0000-0003-3965-6931]{Jie Lin}
\affiliation{ 
Department of Astronomy, University of Science and Technology of China, Hefei, 230026, China; sfhuang999@ustc.edu.cn;twang@ustc.edu.cn; jnac@ustc.edu.cn}
\affiliation{School of Astronomy and Space Sciences,
University of Science and Technology of China, Hefei, 230026, China}

\author[0000-0002-9092-0593]{Ji-an Jiang}
\affiliation{ 
Department of Astronomy, University of Science and Technology of China, Hefei, 230026, China; sfhuang999@ustc.edu.cn;twang@ustc.edu.cn; jnac@ustc.edu.cn}
\affiliation{School of Astronomy and Space Sciences,
University of Science and Technology of China, Hefei, 230026, China}
\affiliation{National Astronomical Observatory of Japan, National Institutes of Natural Sciences, Tokyo 181-8588, Japan}

\author[0000-0002-8482-8993]{Kenta Taguchi}
\affiliation{Okayama Observatory, Kyoto University, 3037-5 Honjo, Kamogata-cho, Asakuchi, Okayama 719-0232, Japan}

\author[0000-0003-2611-7269]{Keiichi Maeda}
\affiliation{Department of Astronomy, Kyoto University, Kitashirakawa-Oiwake-cho, Sakyo-ku, Kyoto, 606-8502, Japan}

\begin{abstract}
Quasi-periodic eruptions (QPEs) and recurrent nuclear transients (RNTs) exhibit recurrent high-amplitude flares from galactic nuclei, yet their characteristic timescales remain poorly understood. In this work, we compile a sample of these systems and investigate empirical scaling relations between flare timescales and black hole masses. We find that the recurrence timescale exhibits a positive but highly scattered dependence on black hole mass in the combined QPE and RNT sample, approximately following $t_{\rm rec}\propto M_{\rm BH}^{1.19^{+0.52}_{-0.47}}$ with an intrinsic scatter of 0.95 dex. Remarkably, we uncover a tight nearly linear relation between recurrence time and flare timescale for QPEs and RNTs, described by $t_{\rm rec}\propto t_{\rm rise}^{1.00\pm0.07}$ with an intrinsic scatter of 0.29 dex. This relation extends across timescales from hours for QPEs to months and years for nuclear transients. We further find that QPEs and RNTs approximately follow a common empirical relation between recurrence time and flare rise time, although the physical origin of this relation remains uncertain. Our findings reveal a common phenomenological timescale link across RNTs, providing a practical framework for characterizing their temporal behavior. 
\end{abstract}

\keywords{Tidal disruption (1696) --- Supermassive black holes (1663) --- Black hole physics (159) --- Accretion (14)}

\section{Introduction} \label{sec:intro}
Quasi-periodic eruptions (QPEs) are a newly discovered, rare phenomenon associated to supermassive black holes (SMBHs) in galactic nuclei \citep{Wevers2026}. They are characterized by extremely high-amplitude flares recurring every few hours, with peak luminosities up to $\sim100$ times of their quiescent level. To date, QPEs have only been identified in a small number of galaxies, including GSN~069 \citep{Shu2018,Miniutti2019}, RXJ1301 \citep{Sun2013,Giustini2020}, eRO-QPE1, eRO-QPE2 \citep{Arcodia2021}, eRO-QPE3, eRO-QPE4 \citep{Arcodia2024}, eRO-QPE5 \citep{Arcodia2025}, XMMSL1~J0249 \citep{Chakraborty2021}, AT~2019vcb \citep{Quintin2023}, AT~2019qiz \citep{Nicholl2024}, AT~2022upj \citep{Chakraborty2025}, ZTF19acnskyy \citep[Ansky,][]{Hernandez-Garcia2025,Chakraborty2025}, candidate 4XMM~J235440.7-373019 \citep{Khan2025} and eRASSt~J234402-352640 \citep[J2344,][]{Baldini2026}. Swift~J0230 is a repeated nuclear transient (RNT) near the center of its host galaxy, and its recurrent X-ray flares make it a long-timescale analog of QPEs with a recurrence period of about a month \citep{Evans2023b,Guolo2024}. In addition, another group of recurrent nuclear transients with recurrence intervals from months to years has recently emerged. Several of these systems have been proposed as candidates for repeated partial tidal disruption event (rpTDE), in which a surviving stellar core is repeatedly disrupted by the central SMBH during successive pericenter passages. Examples include AT~2018fyk \citep{Wevers2023,Pasham2024}, eRASSt~J045650.3-203750 \citep{Liu2023,Liu2024}, RX~J133157.6-324319.7 \citep{Malyali2023}, ASASSN-14ko \citep{Payne2021,Payne2022,Payne2023,Tucker2021}, IRAS~F01004-2237 \citep{Sun2024}, AT~2020vdq \citep{Somalwar2025,Bandopadhyay2024}, AT~2022dbl \citep{Lin2024,Hinkle2024,Zhong2025}, AT~2019aalc \citep{Sniegowska2025,Veres2026}, AT~2021aeuk \citep{Bao2024,Sun2025}, AT~2023adr \citep{Quintin2025,Angus2026,Lanza2026}, AT~2023uqm \citep{Wang2025}, AT~2019azh,  AT~2024pvu \citep{Yao2026,Langis2026}, AT~2022sxl \citep{Ji2025}, AT~2021acak \citep{li2023}, XMMSL2~140446.9-251135 \citep{Saxton2025} and IGR~J12580+0134 \citep{Ma2026}. Some QPE sources, such as GSN~069 \citep{Miniutti2023a,Sheng2021}, have also been suggested to be associated with repeated partial disruption activity. In addition to these sources, another type of recurrent ultra-luminous X-ray (ULX) source near the galaxy NGC~4697 was discovered by \cite{Irwin2016}. Recently, a recurrent ULX XMMU~J122939.7+075333 was reported in the globular cluster RZ~2109 in the Virgo galaxy NGC~4472 \citep{Tiengo2022,Dage2024}. ASASSN-14ko exhibits a periodic outburst every 115 days in the UV/optical bands \citep{Payne2021,Payne2022,Payne2023}, analogous to X-ray QPEs \citep{Huang2023b,Huang2025}. IC~3599 exhibits repeated flares with intervals for tens of years which could be produced by rpTDE \citep{Campana2015} or radiative pressure instability \citep{Grupe1995,Grupe2015,Grupe2026}.

Most QPEs do not show significant UV/optical variability, except for Ansky \citep{Guo2026}. A blackbody component provides a good fit to the X-ray spectra of these eruptions. During QPEs, the black body temperature increases with luminosity, and as a result, the light curves at lower energy lag behind the higher ones \citep{Miniutti2019}. In GSN~069, the QPE signal appears after a tidal disruption event (TDE), and is present only in the low state, while it vanishes in the high state \citep{Shu2018, Sheng2021,Miniutti2023a,Miniutti2023b}. Besides that, some QPEs occur after TDEs, such as AT~2019qiz \citep{Nicholl2024}, AT~2019vcb \citep{Quintin2023,Bykov2025}, AT~2022upj \citep{Chakraborty2025}, Ansky \citep{Zhu2025}, and J2344 \citep{Homan2023}. The spectroscopic analysis indicates that the existence of extended
emission-line regions in the host galaxies of QPEs \citep{Wevers2024,Xiong2025}. These observations suggest a history of activity in the active galactic nucleus (AGN) and TDE produces the disk, providing the environment for the star or black hole to interact with it \citep{Jiang2025}. The circumnuclear environment can be probed through infrared echoes associated with MIR flares \citep{Wu2025,Pasham2025}.

Many scenarios have been proposed to explain the origin of QPEs \citep{Witzany2026}. One scenario involves a white dwarf partially disrupted by SMBH, which can produce QPE \citep{king2020,King2022,King2023,King2023b,Wang2022}. This scenario was previously proposed to explain the physical origin of ULXs \citep{Shen2019}. However, \cite{Metzger2022} suggest that QPEs can also result from material transferred from a binary system orbiting the SMBH. Another possibility is that QPEs are caused by unstable mass transfer from a low-mass star to the SMBH \citep{Linial2023,Lu2023}. Additionally, QPEs can be produced by the instability of the disk with the dominant radiation pressure, which can explain the evolution of the blackbody temperature as a function of luminosity \citep{Sniegowska2020,Pan2021,Pan2022,Pan2023,Pan2025}. The magnetic pressure dominant instability may also trigger QPE signals \citep{Kaur2023}. Furthermore, periodic flares can be induced by the interaction between an object that crosses the disk and the disk itself. This scenario was the first to be applied to explain the 12-yr quasi-periodicity in OJ~287 \citep{Lehto1996,Valtonen2008}, and it may also explain QPEs \citep{Xian2021,Franchini2023,Tagawa2023,Linial2023b,Linial2024b,Zhou2024,Zhou2024b,Zhou2025,Zhou2025b,Jiang2025,HuangX2025,Chen2026}. \cite{Sukova2021} propose that the periodic perturbation of the accretion flow by a star can produce QPE signals. On the other hand, some models suggest that QPEs are related to the self-lensing of a supermassive black hole binary \citep{Ingram2021} or the precession of the inner disk \citep{Raj2021}. However, it remains challenging for these models to reproduce the observed properties of QPEs. For example, self-lensing models do not readily account for several observed properties of QPEs, including their spectral evolution, while disk-instability mechanisms generally struggle to produce the short recurrence and eruption timescales observed around relatively massive black holes. Orbital models (such as star-disk interaction) additionally require a sufficiently long-lived and stable orbital or precessional configuration to account for the quasi-periodic nature of the eruptions \citep{Wevers2026,Guow2026}. 

In this work, we compile a sample of QPEs, RNTs, and rpTDE candidates reported so far, and investigate empirical scaling relations between recurrence timescale, flare rise time, and black hole mass. We show that QPEs and longer-timescale recurrent nuclear transients approximately follow a common timescale relation across several orders of magnitude, suggesting a unified phenomenological framework for recurrent SMBH transients.

In Section~\ref{sec:results}, we present the statistical relations between recurrence time, flare rise time, and black hole mass. In Section~\ref{sec:diss}, we discuss the implications of these empirical relations for different physical models of QPEs and rpTDEs.

\section{Results}\label{sec:results}
\subsection{The role of black hole mass in QPEs and RNTs}
\begin{figure*}[htb]
  \centering
  \subfigure[]{
  \includegraphics[width=0.46\textwidth]{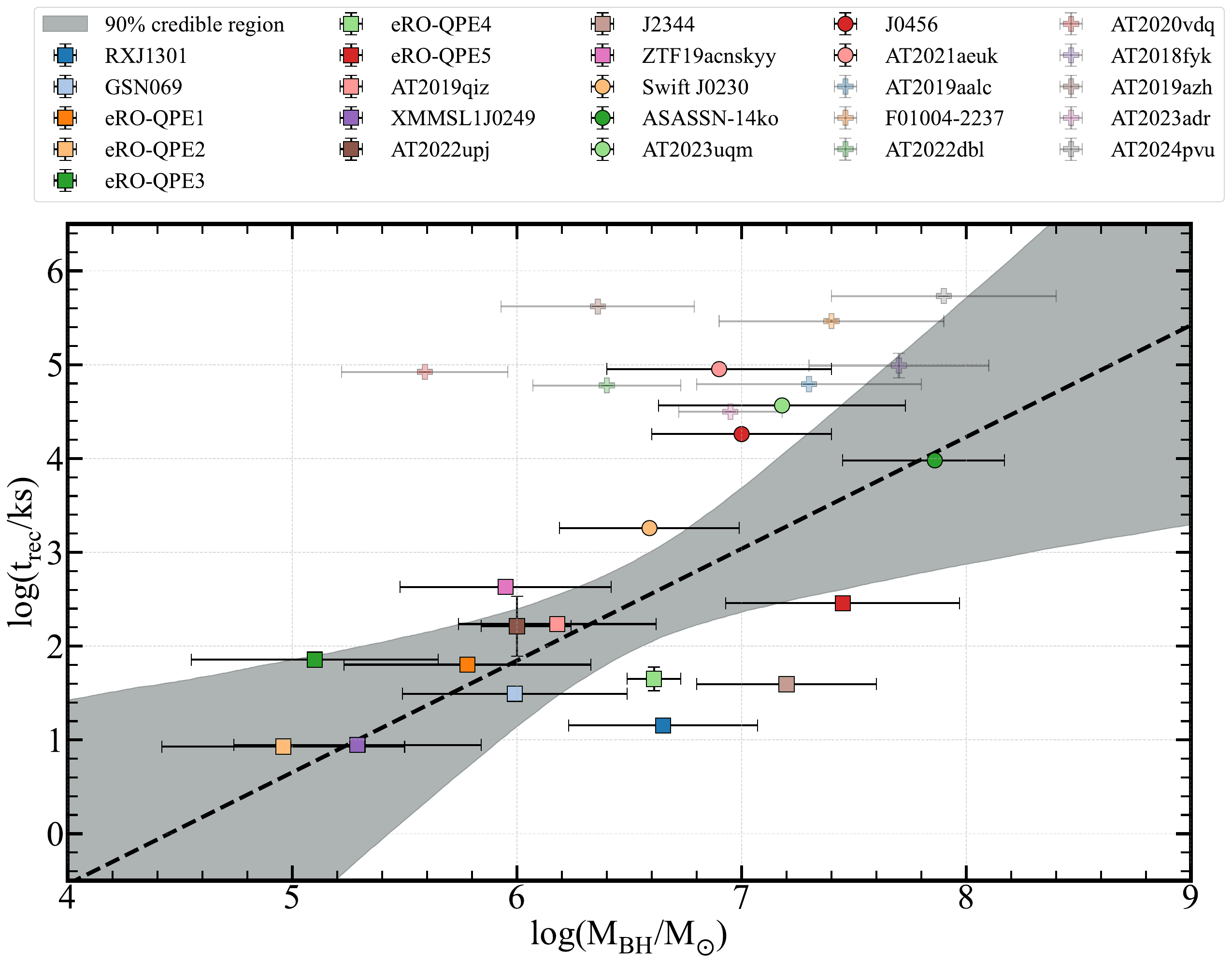}}
  \subfigure[]{\includegraphics[width=0.47\textwidth]{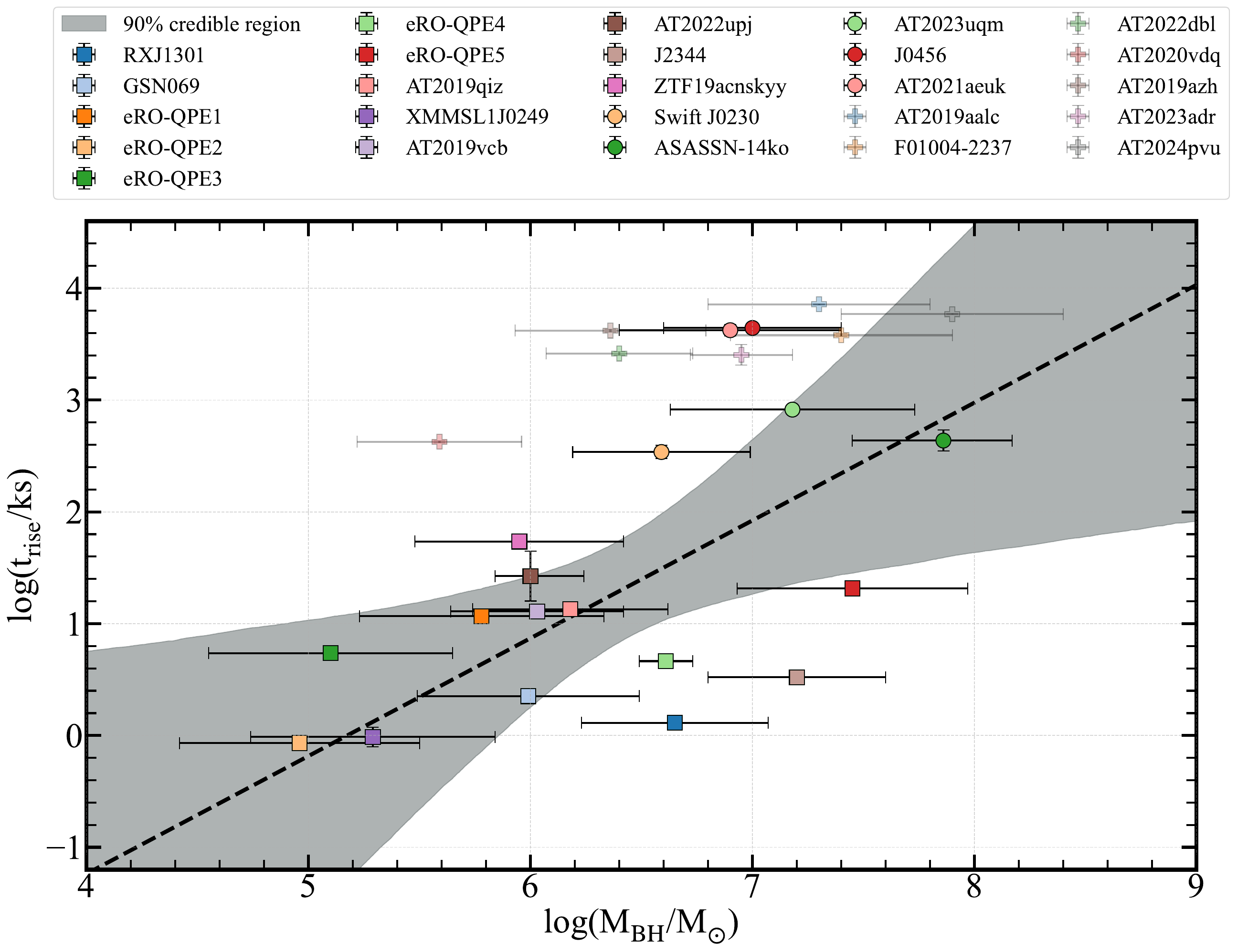}}
  \caption{The relation between the black hole mass and the recurrence time for QPEs, RNTs, and rpTDE candidates. The recurrence times are corrected to the source rest frame. QPEs, RNTs, and rpTDE candidates are shown as squares, circles, and pentagons, respectively. Only QPEs and RNTs are included in the statistical analyses presented in this work. The rpTDE candidates are shown for comparison but are not included in the statistical analysis. The black dashed line represents the median relation from a hierarchical Bayesian linear regression using the QPE and RNT samples, excluding AT~2021aeuk. The regression accounts for measurement uncertainties in both variables. The shaded region represents the 90\% posterior credible region of the regression, derived from the posterior distributions of the slope and intercept and therefore incorporating their uncertainties and covariance.}\label{fig:mbh_time}
\end{figure*}

We collected 26 sources, including QPEs, RNTs, and rpTDE candidates.
Data reduction can be seen in Appendix~\ref{sec:data}. The black hole mass of the sources was derived from the public works of literature. We assumed that the mean separated time between the peaks for these sources was the recurrence time ($\tau_{\rm rec}$) in this work and the rest frame time scale is $t_{\rm rec}=\tau_{\rm rec}/(1+z)$. We fitted the rising part of the light curve in each eruption by a Gaussian function and treated the full width at half maximum as the rise time ($\tau_{\rm rise}$) and corresponding to a rest-frame timescale $t_{\rm rise}=\tau_{\rm rise}/(1+z)$. The details of these parameters are shown in Table \ref{tab:QPE_time}. In this section, we compile the sample of QPEs and RNTs used in our statistical analysis. Only QPEs and RNTs are included in the quantitative analyses and correlation tests presented below. We also include  rpTDEs in Table~\ref{tab:QPE_time} and Figures~\ref{fig:mbh_time} and Figure~\ref{fig:P_rise} for comparison and to provide a broader view of the phenomenological diversity of RNTs; however, these sources are not included in the statistical analyses. AT~2021aeuk is not included in the regression analyses because only two major complete outbursts are available.  

\begin{deluxetable*}{lccccccc}
\tablenum{1}
\tablecaption{Summary of the sources  \label{tab:QPE_time}}
\tablewidth{0pt}
\tabletypesize{\scriptsize}
\tablehead{
\colhead{Name}  & \colhead{Redshift} & \colhead{$\log{(\tau_{\rm rec}/\rm ks)}$} & \colhead{$\log{(\tau_{\rm rise}/\rm ks)}$} &  \colhead{ $\log{\left(\frac{M_{\rm BH}}{M_\sun}\right)}$} & \colhead{Type} & \colhead{Flares number}
& \colhead{References} }
\decimalcolnumbers
\startdata
RXJ1301 & 0.02374 & $1.16\pm 0.001$ & $0.12\pm 0.01$ & $6.65\pm 0.42$ & QPE & 9 & \cite{Wevers2022,Giustini2024} \\
GSN~069 & 0.0181 & $1.50\pm 0.02$ & $0.36\pm0.01$ & $5.99\pm0.5$ & QPE & 5 & \cite{Miniutti2019,Wevers2022} \\
eRO-QPE1 & 0.0505 & $1.82\pm 0.06$ & $1.09\pm0.01$ & $5.78\pm0.55$ & QPE & 15 & \cite{Arcodia2021,Wevers2022} \\
eRO-QPE2 & 0.0175 & $0.94\pm 0.02$ & $-0.06\pm0.02$ & $4.96\pm0.54$ & QPE & 9 & \cite{Arcodia2024b,Wevers2022} \\
eRO-QPE3 & 0.024 & $1.87\pm 0.01$ & $0.75\pm0.03$ & $5.1\pm{0.55}$ & QPE & 2 & \cite{Arcodia2024,Wevers2024}\\
eRO-QPE4 & 0.044 & $1.67\pm 0.13$ &  $0.69\pm0.01$ & $6.61\pm{0.12}$ & QPE & 3 & \cite{Arcodia2024}\\
eRO-QPE5 & 0.1155 & $2.50\pm{0.001}$ & $1.36\pm{0.02}$ & $7.45\pm{0.52}$ & QPE & 3 & \cite{Arcodia2025} \\
XMMSL1J0249	& 0.0186 & $0.95\pm	0.02$ &	$-0.005\pm0.09$ & $5.29\pm0.55$ & QPE & 1.5 & \cite{Chakraborty2021,Wevers2022} \\
AT~2019vcb & 0.089 & - &	$1.15\pm0.03$ & $6.03\pm0.39$ & QPE & - &\cite{Quintin2023,Yao2023}\\
AT~2019qiz & 0.0151 & $2.24\pm{0.003}$ & $1.14\pm{0.003}$ & $6.18\pm{0.44}$ & QPE & 8 & \cite{Nicholl2020,Nicholl2024}\\ 
AT~2022upj & 0.0554 & $2.24\pm{0.32}$ & $1.45\pm{0.22}$ & $6^{+0.24}_{-0.16}$ & QPE & 8 & \cite{Newsome2024,Chakraborty2025} \\
Ansky & 0.024 & $2.64\pm{0.08}$ & $1.75\pm{0.02}$ & $5.95\pm{0.47}$ & QPE & 12 & \cite{Hernandez-Garcia2025,Sanchez-Saez2026} \\
J2344 & 0.1 & $1.64\pm{0.04}$ & $0.56\pm{0.05}$ & $7.2\pm{0.2}$ & QPE & 3 &\cite{Baldini2026,Malyali2026} \\
\hline
Swift~J0230 & 0.03997 & $3.27\pm0.02$ & $2.55\pm0.06$ & $6.6\pm0.4$ & RNT & 11 & \cite{Evans2023b,Guolo2024}\\
ASASSN-14ko & 0.042 & $4.00\pm0.005$ & $2.66\pm0.09$ & $7.86^{+0.31}_{-0.41}$ & RNT & 21 & \cite{Payne2021}\\
J0456 & 0.077 & $4.29\pm0.004$ & $3.68\pm0.02$ & $7.4\pm0.5$ & RNT & 5 & \cite{LiuZ2023}\\ 
AT~2021aeuk & 0.2336 & $5.04\pm{0.001}$ & $3.72\pm{0.05}$ & $6.9\pm{0.5}$ & RNT & 2 & \cite{Bao2024,Sun2025}\\
AT~2023uqm & 0.238 & $4.66\pm{0.001}$ & $3.01\pm{0.02}$ & $7.18\pm{0.55}$ & RNT & 5 & \cite{Wang2025} \\
\hline
AT~2018fyk & 0.059 & $5.02\pm0.13$ & - & $7.7\pm0.4$ & rpTDE & 2 & \cite{Wevers2023} \\
AT~2019aalc & 0.03557 & $4.81\pm{0.0002}$ & $3.87\pm{0.004}$ & $7.3\pm{0.5}$ & rpTDE & 2 & \cite{Sniegowska2025,Veres2026}\\
AT~2019azh & 0.022 & $5.63\pm{0.001}$ & $3.63\pm{0.01}$ & $6.36^{+0.43}_{-0.40}$ & rpTDE & 2 & \cite{Wevers2020,Yao2026} \\
AT~2020vdq & 0.045 & $4.92\pm{0.0004}$ & $2.63\pm{0.009}$ & $5.59\pm{0.37}$ & rpTDE & 2 & \cite{Somalwar2025} \\
AT~2022dbl & 0.0284 & $4.79\pm{0.0003}$ & $3.28\pm{0.02}$ & $6.4\pm{0.33}$ & rpTDE & 2 & \cite{Lin2024} \\
AT~2023adr & 0.131 & $4.55\pm{0.003}$ & $3.46\pm{0.09}$ & $6.95\pm{0.23}$ & rpTDE &  2 & \cite{Quintin2025, Angus2026} \\
AT~2024pvu & 0.048 & $5.75\pm{0.0004}$ & $3.79\pm{0.005}$ & $7.90\pm{0.5}$ & rpTDE & 2 & \cite{Langis2026,Yao2026} \\
F01004-2237 & 0.1178 & $5.51\pm{0.01}$ & $3.63\pm{0.13}$ & $7.4\pm{0.5}$ & rpTDE & 2 & \cite{Dou2017,Sun2024} \\
\enddata
\tablecomments{
Columns (2)--(5) list the redshift, recurrence time, rise time, and black hole mass, respectively. Column (7) gives the number of flares used to determine $t_{\rm rec}$ in this work. For AT~2019vcb, only the rising phase is detected; therefore, the QPE recurrence time cannot be determined. The black hole masses of RXJ1301, GSN~069, eRO-QPE1, eRO-QPE2, and XMMSL1J0249 were adopted from \cite{Wevers2022}, while those of AT~2019vcb, AT~2019qiz, Ansky, Swift~J0230, and J0456 were adopted from \cite{Yao2023,Nicholl2020,Sanchez-Saez2026,Guolo2024,LiuZ2023}, respectively. These values were estimated using the $M_{\rm BH}$--$\sigma$ relation, as was the black hole mass of eRO-QPE3 \citep{Wevers2024}. The black hole mass of AT~2022upj, together with other relevant source information, was adopted from \cite{Newsome2024}, where it was estimated using the $M_{\rm BH}$--$M_{\rm bulge}$ relation. The black hole masses of J2344 \citep{Malyali2026}, ASASSN-14ko \citep{Payne2021}, AT~2019aalc \citep{Sniegowska2025}, and AT~2021aeuk \citep{Sun2025} were estimated using broad emission-line scaling relations. For eRO-QPE4, we adopted the stellar mass reported by \cite{Arcodia2024} and estimated its black hole mass using Equations (4) and (5) of \cite{Reines2015}. The black hole mass of eRO-QPE5 was adopted from \cite{Arcodia2025}, where it was estimated using the $M_*$--$M_{\rm BH}$ relation. The black hole mass of AT~2023uqm was likewise estimated using the $M_*$--$M_{\rm BH}$ relation \citep{Wang2025}. For the remaining rpTDEs, the black hole masses were estimated using the $M_{\rm BH}$--$\sigma$ relation. Uncertainties are propagated from the uncertainties of the fitted peak times used to determine the recurrence intervals. They therefore represent the measurement uncertainties of the individual recurrence times and do not include any intrinsic scatter or population-level variation in recurrence times. For AT~2021aeuk, we measure the timescale using only the two complete flares observed in 2019 and 2023.}
\end{deluxetable*}

We fitted the correlations using the \texttt{linmix} package \citep{Kelly2007}, which implements a hierarchical Bayesian linear regression method that accounts for measurement uncertainties in both variables and intrinsic scatter. The fitting was performed in logarithmic space using a linear model of the form
$\log y = h + a \log x$.
The quoted uncertainties of the fitting parameters represent the 1$\sigma$ credible intervals derived from the posterior distributions.

Here, we fitted the data with a function of $\log t=h + a\log M_{\rm BH}$ for both the relation between $t_{\rm rec}$ vs $M_{\rm BH}$ and $t_{\rm rise}$ vs $M_{\rm BH}$. For the $t_{\rm rec}-M_{\rm BH}$ relation of in QPEs and RNTs, we obtained
\begin{equation}
    \log (t_{\rm rec}/\rm ks)=(-5.3^{+3.01}_{-3.45})+(1.19^{+0.52}_{-0.47})\log\left(\frac{M_{\rm BH}}{M_\sun}\right),
\end{equation}
with an intrinsic scatter of 0.95 dex.
Using the python package \texttt{scipy.stats.spearmanr}, we obtain the Spearman rank correlation coefficient of 0.53 and the P-value of 0.02, respectively. Through \texttt{scipy.stats.pearsonr}, we derive the Pearson correlation coefficient of 0.63 and the P-value of 0.008, respectively.

The relation between black hole mass and timescale is shown in Figure~\ref{fig:mbh_time}. The combined QPE and RNT sample shows a positive association between recurrence time and black hole mass, approximately described by $t_{\rm rec}\propto M_{\rm BH}^{1.19}$. However, the correlation is weak when QPEs and RNTs are considered separately, suggesting that the correlation in the combined sample may be partly driven by the different parameter ranges occupied by the two populations.\cite{Webbe2023} obtained a relation of $t_{\rm rec}\propto M_{\rm BH}$ in QPEs, but used a linear function in the fitting that differs from ours. While for the work focusing on reported QPEs by \citep{Zhou2025b}, $t_{\rm rec}\propto M_{\rm BH}^{0.8}$ was derived. 

 We also tested the correlation using the QPEs alone and obtained the Spearman rank correlation coefficient of 0.29 (P-value of 0.18), and the Pearson correlation coefficient is 0.36 (P-value of 0.25). For RNTs alone, the Spearman rank correlation coefficient of 0.40 (P-value of 0.3), and the Pearson correlation coefficient is 0.43 (P-value of 0.57) were obtained. We can see that when QPEs and RNTs are analyzed separately by class, the correlations are weak in both cases, whereas the correlation becomes significantly stronger when the two populations are combined. It is worth noting that the black holes hosting QPEs are generally of relatively low mass and have short recurrence timescales, $t_{\rm rec}$, while the RNTs included in our sample typically host more massive black holes than most QPEs and exhibit longer $t_{\rm rec}$. Therefore, combining these two populations could potentially introduce an artificial correlation. However, an important factor that should also be considered is the relatively large uncertainty in the black hole mass measurements, which may significantly affect the inferred correlation. After applying dynamical corrections to the black hole masses of QPEs, \cite{Zhou2025b} found a significant correlation between $M_{\rm BH}$ and $t_{\rm rec}$, highlighting the importance of accurate black hole mass measurements. Therefore, we still have reason to believe that, with improved precision in black hole mass measurements, an intrinsic correlation between $t_{\rm rec}$ and black hole mass may exist, at least for QPEs. For RNTs, if the two phenomena share a similar dynamical origin, a similar correlation would also be expected.

While for the relation $t_{\rm rise}-M_{\rm BH}$ of QPEs and RNTs, we obtained the fitted results as
\begin{equation}
    \log (t_{\rm rise}/\rm ks)=(-5.38^{+2.95}_{-3.33})+(1.04^{+0.51}_{-0.46})\log\left(\frac{M_{\rm BH}}{M_\sun}\right),
\end{equation}
with an intrinsic scatter of 1 dex.
Here we derived the Spearman rank correlation coefficient of 0.50 and the P-value of 0.02, respectively. Furthermore, we obtain the Pearson correlation coefficient of 0.58 and the P-value of 0.02, respectively. 

The rise timescale of QPEs and RNTs also shows a positive dependence on black hole mass, approximately following $t_{\rm rise}\propto M_{\rm BH}^{1.04}$, although the relation has substantial intrinsic scatter. The relation is very different from the case in TDEs, where
$t_{\rm rise}\propto M_{\rm BH}^{1/2}$ is observed in found TDEs \citep{Velzen2019,Velzen2020,Huang2023}. Our result suggests that the physical mechanism of QPEs may have a different origin from that of TDEs.

\subsection{The $t_{\rm rec}$ and $t_{\rm rise}$ in QPEs and RNTs}

\begin{figure*}[htb]
  \centering
  \includegraphics[width=0.7\textwidth]{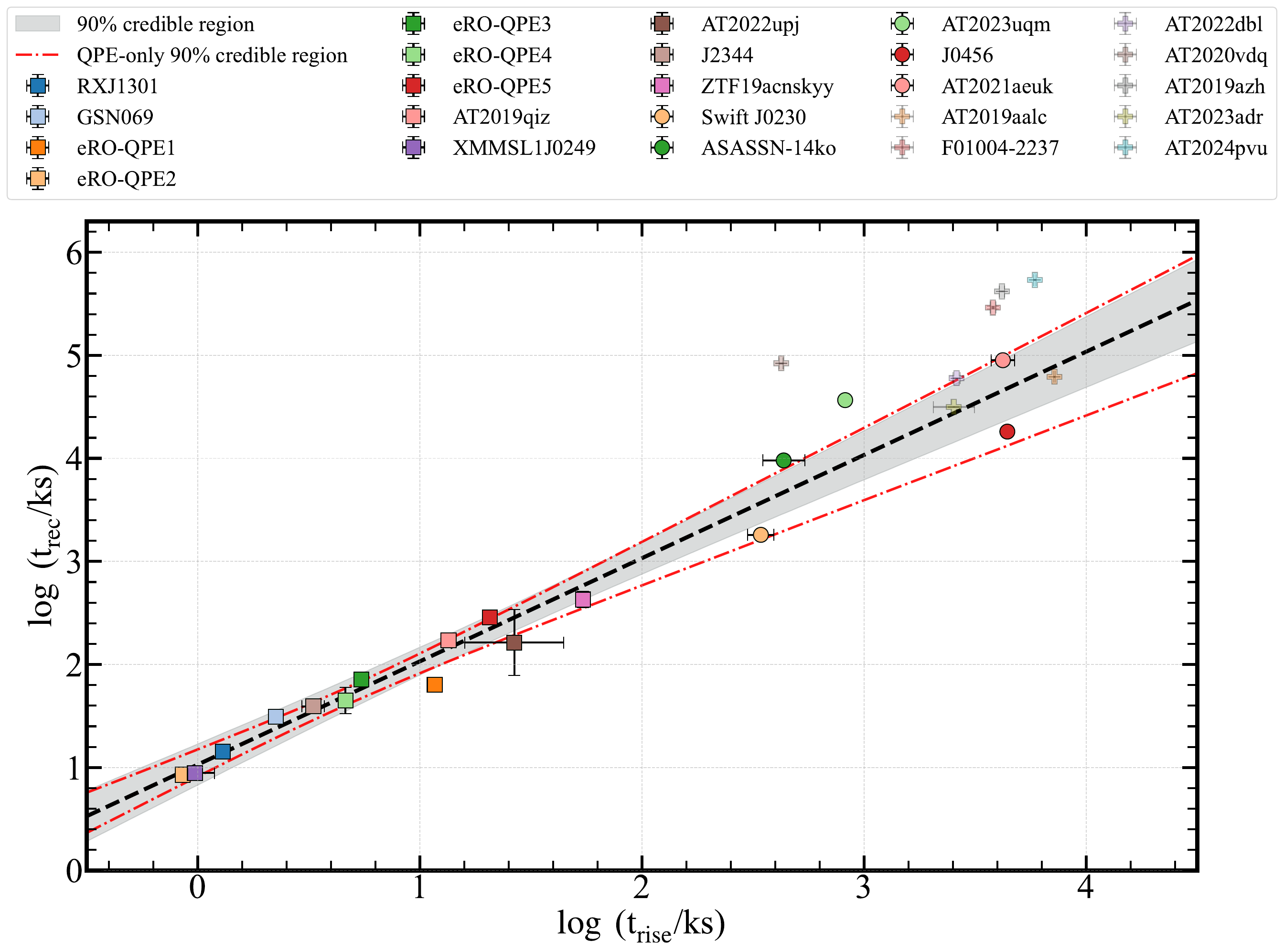}
  \caption{The relation between the rise time and recurrence time for QPEs, RNTs, and rpTDE candidates. The statistical analysis includes only QPEs and RNTs. The rpTDE candidates are shown for comparison and to illustrate their location relative to the QPE and RNT populations, but they are not included in the correlation analysis. The dashed black line shows the best-fit relation obtained from a hierarchical Bayesian regression that accounts for uncertainties in both variables. The shaded region represents the 90\% posterior credible region of the regression, derived from the posterior distributions of the slope and intercept and therefore incorporating their uncertainties and covariance. Squares denote QPEs, circles denote RNTs with at least three observed outbursts, and crosses denote RNTs with only two complete outbursts and rpTDE candidates. The red dash-dotted lines mark the 90\% posterior credible region inferred from a fit to the QPE-only sample, incorporating the posterior uncertainties of the regression parameters as well as the intrinsic scatter.}\label{fig:P_rise}
\end{figure*}

The correlation between $t_{\rm rec}$ and $t_{\rm rise}$ is shown in Figure \ref{fig:P_rise}. It should be noted that $t_{\rm rec}$ exhibited a significant correlation with $t_{\rm rise}$. Then we fitted the data points with the function  $\log t_{\rm rec}=h + a\log t_{\rm rise}$.

We fitted the data of QPEs and obtained the results 
\begin{equation}
    \log (t_{\rm rec}/\rm ks)=(1.05\pm 0.07)+(0.97^{+0.08}_{-0.09})\log (t_{\rm rise}/\rm ks),
    \label{eq:qpe}
\end{equation}
with an intrinsic scatter of 0.14 dex.
For this case, the Spearman rank correlation coefficient of 0.97 and the  P-value $\ll 0.01$, respectively. Besides, we obtained the Pearson correlation coefficient of 0.97 and the P-value $\ll 0.01$, respectively. 

In order to check the reliability, we fitted the data of QPEs and RNTs together. Due to only two complete flares in AT~2021aeuk, we did not include this source in the fitting. A result of 
\begin{equation}
    \log (t_{\rm rec}/\rm ks)=(1.03\pm 0.12)+(1.00\pm0.07)\log (t_{\rm rise}/\rm ks)
    \label{eq:qpe_and_rnt}
\end{equation}
with 0.29 dex intrinsic scatter. We obtained the Spearman rank correlation coefficient of 0.99 and the P-value $\ll 0.01$, respectively. Besides, we obtained the Pearson correlation coefficient of 0.98 and the P-value $\ll 0.01$, respectively. However, when we only fitted RNTs and rpTDEs, large scattering occurred and a weak correlation was derived. 

A similar relation in QPEs is shown by \cite{Guolo2024,Arcodia2024,Nicholl2024,Chakraborty2025,Hernandez-Garcia2025,Hernandez2025b} and our results indicate that this relation may also extend to RNTs. Through this relation, it is possible to predict the recurrence time based on the rise time scale of the flare.

\section{Discussion}\label{sec:diss}

The origin of QPEs remains uncertain, and several physical scenarios have been proposed to explain their recurrent eruptions. In this section, we discuss representative models for QPEs and examine whether the observed scaling relations between recurrence time, rise time, and black hole mass can provide constraints on their physical origin. We further compare QPEs with longer-timescale RNTs, and explore whether these systems may belong to a broader family of recurrent SMBH transients.

\subsection{Implications for the origin of QPEs}

The possible dependence of QPE characteristic timescales on black hole mass may provide clues to the physical origin of these eruptions. Although the current QPE sample does not provide statistically significant evidence for a correlation between the characteristic timescales and $M_{\rm BH}$, the observed trends may still motivate a physical interpretation in terms of processes operating near the SMBH.

One promising interpretation is that QPEs are powered by repeated interactions between an orbiting compact object and the accretion disk surrounding the SMBH \citep{Xian2021,Franchini2023,Linial2023,Linial2024,Linial2024b,Tagawa2023,Zhou2024,Zhou2024b,Zhou2025,Jiang2025,Liu2026}. In this picture, a star or stellar-mass black hole periodically crosses the disk and generates shocks that produce luminous X-ray flares. This scenario is conceptually similar to the model proposed for OJ~287, where the secondary SMBH impacts the accretion disk of the primary and produces quasi-periodic optical outbursts \citep{Lehto1996,Valtonen2008,Valtonen2016,Laine2020}.

The recurrence time is naturally associated with the orbital period of the impactor,
\begin{equation}
P=2\pi \sqrt{\frac{r_a^3}{GM_{\rm BH}}},
\end{equation}
where $r_a$ is the semimajor axis of the orbit. If the impact occurs at a characteristic radius in units of gravitational radius, the recurrence timescale approximately follows
\begin{equation}
t_{\rm rec}\propto M_{\rm BH}.
\end{equation}

For the QPEs alone, however, the current sample does not provide statistically significant evidence for a correlation between $t_{\rm rec}$ and $M_{\rm BH}$. The positive trend found in the combined QPE and RNT sample is therefore not taken as evidence for a QPE-specific correlation. Nevertheless, the approximate scaling expected in the star--disk interaction scenario provides a useful theoretical comparison for future, better constrained QPE samples. Hydrodynamic calculations further suggest that the shocked gas produced after impact expands with a velocity related to the local Keplerian speed, naturally linking the flare evolution to the orbital dynamics of the system \citep{Pihajoki2016,Franchini2023}. Previous theoretical work has also predicted that the flare rise time may scale approximately linearly with black hole mass \citep{Linial2023b}. This provides a qualitative theoretical expectation that can be tested with larger QPE samples, although the current observations do not yet provide statistically significant evidence for such a relation.

The impact model may also qualitatively explain the observed spectral evolution of QPEs. After the collision, an expanding shocked gas cloud forms, and radiation can escape efficiently once the ejecta become optically thin \citep{Valtonen2016}. However, several issues remain unresolved. Recent observations of GSN~069 show that the QPE activity disappears when the quiescent luminosity rises above a critical level \citep{Miniutti2023b}, implying that the accretion-flow state strongly influences the production of eruptions. In addition, the observed temperatures and radiated energies of QPEs are difficult to reconcile with impacts from ordinary main-sequence stars, while compact impact objects such as stellar-mass black holes remain viable \citep{Guow2026,Liu2026}. The approximately linear relation between flare rise time and recurrence time may be broadly consistent with a dynamical origin, although its physical interpretation remains uncertain.

Alternative scenarios have also been proposed. Radiation-pressure-driven instabilities in the inner accretion disk can generate recurrent eruptions resembling QPEs \citep{1976MNRAS.175..613S,1974ApJ...187L...1L}. In magnetically supported disks, the unstable region may become sufficiently compact to reproduce QPE-like timescales \citep{Pan2021,2021ApJ...916...61F,Kaur2023,Chashkina2024}. Numerical calculations by \cite{Pan2022} further showed that the recurrence and rise times in disk-instability models can approximately scale with black hole mass, broadly consistent with the observed trends reported in this work. However, disk-instability models may face difficulties in explaining several timing properties of QPEs. 

Overall, the observed timescale--mass trends are broadly consistent with models in which the characteristic variability timescales are related to dynamical processes near the SMBH. In particular, repeated star--disk or compact-object--disk interactions provide a plausible explanation for the observed recurrence and rise-time scaling relations, as the recurrence timescale can naturally be linked to the orbital motion of the impacting object. However, given the limited sample size and the lack of statistically significant correlations with $M_{\rm BH}$, these relations alone cannot uniquely distinguish the star--disk interaction scenario from alternative models.

\subsection{A possible connection between QPEs and RNTs}

An increasing number of RNT and rpTDE candidates have been discovered in recent years. These systems exhibit recurrence timescales ranging from months to years, substantially longer than those of QPEs. Nevertheless, despite the large difference in characteristic timescales, our results show that QPEs and RNTs approximately follow the same empirical relation between recurrence time and flare rise time.

This continuity may suggest that at least some of these systems share similar dynamical ingredients. In both QPEs and RNTs systems, repeated flares may ultimately be associated with bound objects orbiting SMBHs. In QPE models based on star--disk or compact-object--disk interactions, recurrent flares are naturally linked to repeated passages through the accretion flow. Similarly,  in RNT systems, including ASASSN-14ko, J0456, AT~2018fyk, and AT~2023uqm, have been interpreted as rpTDEs, a surviving stellar core can undergo repeated partial stripping during successive pericenter passages. In both cases, the recurrence timescale may be related to orbital motion around the SMBH, although the mapping between the observed flare timescale and the underlying orbital dynamics may differ between the two classes.

\begin{figure*}[htbp]
   \centering
   \includegraphics[width=0.9\textwidth]{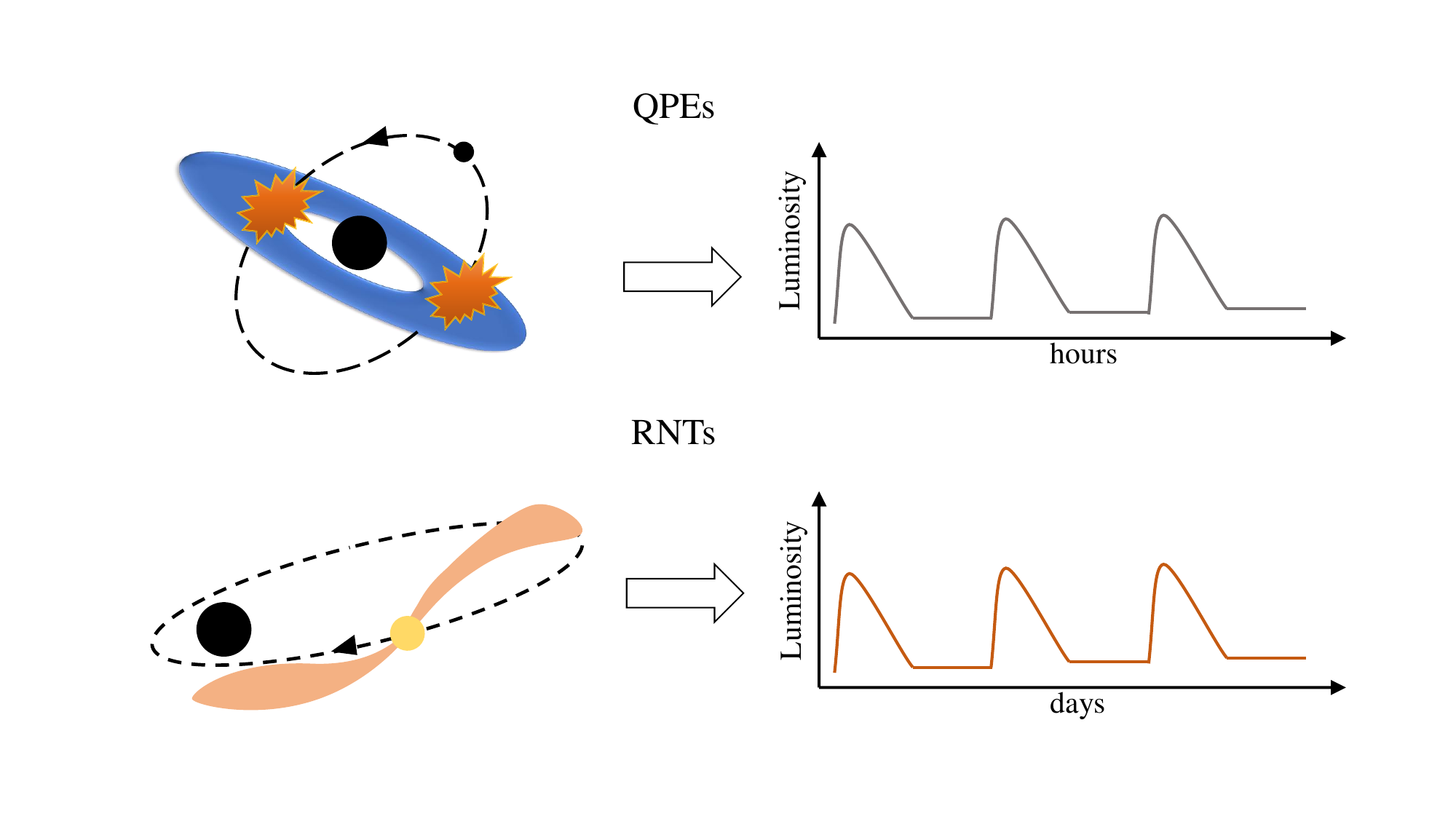}
   \caption{Schematic diagram illustrating a possible connection between QPEs and RNTs systems. QPEs exhibit short recurrence timescales, typically ranging from hours to days, while RNTs show much longer recurrence timescales of months to years. In both cases, recurrent flares may be associated with repeated orbital interactions around SMBHs. The luminosity and time axes are schematic and not to scale.}
   \label{fig:model}
\end{figure*}

Figure~\ref{fig:model} illustrates a schematic picture connecting these populations. Within this phenomenological framework, QPEs may represent the short-timescale end of recurrent SMBH transients, while RNTs occupy the long-timescale regime. The observed scaling relation extending across several orders of magnitude in timescale may therefore indicate a broader family of recurrent nuclear transients than previously recognized. The dynamical connection between QPEs and rpTDEs has also been discussed in \citet{Hinkle2026}. However, the similarity of the empirical timescale relation does not necessarily imply a common emission mechanism. If optical RNTs are powered by a mechanism distinct from that responsible for QPEs, such as repeated stripping and subsequent accretion of stellar material, there is no a priori reason for their rise times to scale with $M_{\rm BH}$ or $t_{\rm rec}$ in the same manner as those of QPEs. Conversely, if RNTs are produced by a mechanism similar to the star--disk interaction scenario proposed for QPEs, the predominance of optical rather than X-ray emission in these systems requires additional explanation. Thus, the physical origin of the apparent continuity between QPEs and RNTs remains an open question.

The empirical relation between $t_{\rm rec}$ and $t_{\rm rise}$ provides a potential predictive framework for estimating approximate recurrence timescales in candidate RNT systems. Since the sources in our sample exhibit multiple recurrent flares and follow the same relation as QPEs, candidate systems with a measured flare rise time may have their recurrence timescales approximately inferred from this relation. For example, because only part of the flare evolution of AT~2019vcb was captured \citep{Quintin2023}, Equation~\ref{eq:qpe} predicts a recurrence timescale of approximately 1.4 days. Similarly, using the rise time reported for AT~2021mhg by \cite{Yao2023}, we estimate an expected recurrence timescale of $\sim$400 days. Interestingly, another optical flare detected by ZTF occurred close to this predicted epoch \citep{Somalwar2025}, although the nature of this flare and its association with the nuclear transient remain uncertain. More recently, AT~2021aeuk exhibited a new optical flare whose occurrence time is also consistent with the predicted recurrence window \citep{WangY2026}.

As shown in Figure~\ref{fig:P_rise}, most of QPEs are consistent with the $t_{\rm rec}$--$t_{\rm rise}$ relation within the 90\% credible region, and the majority of RNTs also follow this relation. Therefore, two-flare sources located within this region, including AT~2019aalc, AT~2022dbl, and AT~2023adr, represent promising targets for future monitoring. If these systems follow the same empirical timescale relation, future outbursts may occur within the recurrence windows inferred from their observed rise times. Notably, a third flare from AT~2022dbl has not yet been detected \citep{Hinkle2026}, which may reflect either incomplete temporal coverage or intrinsic diversity among rpTDEs.

The physical origin of the rise and duration timescales in QPEs remains poorly understood. Nevertheless, similar timescale--mass correlations have been explored in other SMBH accretion-powered transients. For example, in the SMBH binary candidate OJ~287, which exhibits 12-year quasi-periodic double-peaked optical outbursts, the outburst duration has been proposed to be associated with the bremsstrahlung cooling timescale, resulting in $t_{\rm outburst}\propto M_{\rm BH}^{20/21}$ \citep{Pihajoki2016}. Similarly, theoretical and observational studies of TDEs have suggested a possible dependence of the characteristic rise timescale on black hole mass, approximately following $t_{\rm rise}\propto M_{\rm BH}^{1/2}$ \citep{Lodato2011,Velzen2019,Velzen2020,Huang2023,Yao2023}. However, the emission processes in TDEs are complex and can be strongly affected by radiative transfer, viewing geometry, and the circumnuclear environment. Therefore, the physical origin of the observed $t_{\rm rec}$--$t_{\rm rise}$ relation remains uncertain.

Nevertheless, the current observations do not establish that QPEs and RNTs originate from a single physical mechanism. The observed continuity may instead reflect a common dependence on underlying dynamical timescales rather than identical emission processes. In addition, the observed flare separation in TDE-like systems may not directly trace the orbital period, because stream collisions, circularization processes, and radiation reprocessing can substantially modify the observed light curves \citep{Piran2015,Roth2016,Metzger2016,Lu2020}. Therefore, the schematic picture shown in Figure~\ref{fig:model} should be interpreted as a phenomenological illustration of the observed scaling relations rather than a unique physical model. Thus, the observed $t_{\rm rec}$--$t_{\rm rise}$ relation should be regarded primarily as an empirical phenomenological connection at present, rather than evidence for a unified emission mechanism.

\section{Conclusion}

We have investigated empirical scaling relations between recurrence times, flare timescales, and black hole masses in QPEs and RNTs.

We find a positive trend between recurrence time and black hole mass in the combined QPE and RNT sample, 
\begin{equation}
t_{\rm rec}\propto M_{\rm BH}^{1.19^{+0.52}_{-0.47}},
\end{equation}
although the relation has substantial scatter, and the corresponding correlations within the individual QPE and RNT populations are not statistically significant.

More significantly, we identify a tight empirical relation between recurrence time and flare timescale,
\begin{equation}
t_{\rm rec}\propto t_{\rm rise}^{1.00\pm0.07},
\end{equation}
which holds across a wide range of timescales spanning QPEs and longer-duration recurrent nuclear transients. This relation suggests the existence of a common phenomenological connection linking the temporal properties of these systems.

Our results indicate that recurrent nuclear transients may share a unified empirical timescale framework, despite differences in their physical origins or triggering mechanisms. These scaling relations provide a practical tool for characterizing and predicting recurrence behavior in newly discovered systems.

The current sample remains limited and heterogeneous, and the observed correlations may be affected by selection effects, cadence biases, and uncertainties in defining characteristic flare timescales. Future time-domain surveys will be essential to test the robustness of these empirical relations and to determine whether they represent a true physical universality or a phenomenological coincidence. These scaling relations also provide a first-order empirical estimator for recurrence timescales in RNTs based on their observed flare rise times.

\section*{acknowledgments}

We thank the anonymous referee for providing valuable comments and suggestions, which helped to improve the manuscript.
The authors thank Dr. Joheen Chakraborty for providing the data of AT~2022upj.  The authors thank Jingbo Sun for providing the information of AT~2021aeuk. The authors thank Prof. Zhen Pan for helpful discussions and valuable suggestions. The authors thank Dr. Xin Pan for helping us understand disk instability. The authors also thank Prof. Xinwu Cao for the fruitful discussion and advice. The authors also thank Prof. Hongxing Yin and Dr. Hao Liu for the discussion about X-ray binaries.
This work is supported by Strategic Priority Research Program of the Chinese Academy of Sciences (XDB0550200), National Key Research and Development Program of China (2023YFA1608100), the National Natural Science Foundation of China (grants 12192221,12393814,12522303), and the Fundamental Research Funds for Central Universities (WK2030000097). The authors appreciate the support of the Cyrus Chun Ying Tang Foundations and the Frontier Scientific Research Program of the Deep Space Exploration Laboratory (2022-QYKYJH-HXYF-012). We acknowledge the use of public data from the Swift data archive. The authors thank the Swift ToO team for accepting our proposal and executing the observations. Based on observations obtained with XMM-Newton, an ESA science mission with instruments and contributions directly funded by ESA Member States and NASA. 
This work makes use of data from the Zwicky Transient Facility. Based on observations obtained with the Samuel Oschin Telescope 48-inch and the 60-inch Telescope at the Palomar Observatory as part of the ZTF project. ZTF is supported by the National Science Foundation under Grant No. AST-2034437 and a collaboration including Caltech, IPAC, the Weizmann Institute for Science, the Oskar Klein Center at Stockholm University, the University of Maryland, Deutsches Elektronen-Synchrotron and Humboldt University, the TANGO Consortium of Taiwan, the University of Wisconsin at Milwaukee, Trinity College Dublin, Lawrence Livermore National Laboratory, and IN2P3, France. Operations are conducted by COO, IPAC, and UW. The ZTF forced-photometry service was funded under the Heising-Simons Foundation grant \#12540303 (PI: Graham).

\vspace{5mm}
\facilities{Swift/XRT, Swift/UVOT, XMM-Newton, NICER and ZTF}

\software{astropy \citep{2013A&A...558A..33A}, HEASoft \citep{2014ascl.soft08004N}, Xspec \citep{1996ASPC..101...17A}, Matplotlib \citep{2007CSE.....9...90H}, linmix \citep{Kelly2007}. }

\appendix
\section{Data Reduction} \label{sec:data}
\subsection{Swift X-ray photometry}
We obtained the public data of Swift~J0230, J0456 and AT~2018fyk from the High Energy Astrophysics Science Archive Research Center (HEASARC) website. We processed the \emph{Swift} data with \texttt{HEASoft 6.31.1}. We use the tasks \texttt{xrtpipeline} and \texttt{xrtproducts} to produce the products. We chose the source region as a circular region centered on the object with a radius of $47.1^{\prime\prime}$, and the background region as an annulus centered on the source with the inner and outer radius of $100^{\prime\prime}$ and $200^{\prime\prime}$, respectively.

\subsection{Swift UV/optical photometry}
For ASASSN-14ko, we ran the task \texttt{uvotimsum} to sum up the Swift/UVOT images and performed the photometry by executing \texttt{uvotsource} with the source and background regions as circles with radii of $10^{\prime\prime}$ and $40^{\prime\prime}$, respectively. We used an online tool to derive the Galactic extinction value of $E(B-V)=0.043$ following \cite{Schlafly2011}, and then corrected the magnitude for each band using the extinction law of \cite{Cardelli1989}. Since Swift had a high observation frequency and a high signal-to-noise ratio in the \textsl{UVW1} band for this source, we only used this single band to perform the light curve analysis. We adopted Galactic extinction values of 0.27 for the \textsl{UVW1}.

\subsection{XMM-Newton photometry}
We accessed the XMM-Newton data from the HEASARC website. We conducted a light curve analysis of QPEs, including GSN~069 (ObsID: 0831790701, PI: Giustini), RX~J1301.9-2747 (ObsID: 0864560101, PI: Giustini), eRO-QPE1 (ObsID: 0861910301, PI: Arcodia), eRO-QPE2 (ObsID: 941270201, PI: Arcodia), eRO-QPE3 (ObsID: 0883770101, PI: Arcodia), eRO-QPE4 (ObsID: 0883770401, PI: Arcodia), eRO-QPE5 (ObsID: 0954190401, PI: Arcodia), XMMSL1~J0249 (ObsID: 0411980401, PI: Schartel) and AT~2019cvb (ObsID: 0871190301, PI: Gezari). In preparation, we used the tasks \texttt{cifbuild} and \texttt{odfingest}, and ran \texttt{xmmextractor} to generate the light curves and spectra.

\subsection{ZTF photometry}
The optical data of TDEs from the Zwicky Transient Facility \citep[ZTF;][]{Masci2019,Bellm2019,Graham2019} were retrieved from Forced Photometry Service \cite{Masci2023}.  Based on \cite{Schlafly2011} and utilizing the online tool\footnote{\url{https://irsa.ipac.caltech.edu/applications/DUST/}}, we corrected the Galactic extinction in optical bands.

\bibliography{QPE}{}
\bibliographystyle{aasjournal}

\end{document}